# Mapping the "long tail" of research funding:
# A topic analysis of NSF grant proposals in the Division of Astronomical Sciences

Gretchen R. Stahlman
*Rutgers University, School of Communication & Information*
gretchen.stahlman@rutgers.edu

P. Bryan Heidorn
*University of Arizona, School of Information*
heidorn@email.arizona.edu

**Abstract**

"Long tail" data are considered to be smaller, heterogeneous, researcher-held data, which present unique data management and scholarly communication challenges. These data are presumably concentrated within relatively lower-funded projects due to insufficient resources for curation. To better understand the nature and distribution of long tail data, we examine National Science Foundation (NSF) funding patterns using Latent Dirichlet Analysis (LDA) and bibliographic data. We also introduce the concept of "Topic Investment" to capture differences in topics across funding levels and to illuminate the distribution of funding across topics. This study uses the discipline of astronomy as a case study, overall exploring possible associations between topic, funding level and research output, with implications for research policy and practice. We find that while different topics demonstrate different funding levels and publication patterns, dynamics predicted by the "long tail" theoretical framework presented here can be observed within NSF-funded topics in astronomy.

**Introduction**

Literature across disciplines refers to "long tail" data as generally smaller, heterogeneous, researcher-held data, and both a challenge and opportunity for data management and scholarly communication. Through analysis of National Science Foundation (NSF) funding patterns using Latent Dirichlet Analysis (LDA) and bibliographic data, we shed light on differences between projects with proportionally "large" funding compared to projects with more modest funding, where long tail data are presumably abundant due to insufficient resources for data curation. We also introduce the concept of "Topic Investment" to capture differences in topics across funding levels and to illuminate the distribution of funding across topics. The study presented here uses the discipline of astronomy as a case study, overall exploring possible associations between topic, funding level and research output, with implications for research policy and practice.

As a concept borrowed from economics, the "long tail" originally refers to niche markets, where some consumer goods are in high demand, but where obscure items may attract attention and become useful if readily available to be discovered (Anderson, 2007). In 2008, Heidorn demonstrated a financial distribution of research funded by the National Science Foundation that resembles the power-law distribution of long tail economics, with 20% of funding in the "head" and 80% in the "tail". Long before the introduction of NSF's data management plan requirement for grant proposals in 2009, large NSF projects were required to plan for data management. Heidorn theorized that an abundance of inaccessible "dark data" corresponding to smaller projects are concentrated in the long tail of the funding distribution (2008, 2011) – data that could also become useful to other researchers if adequate resources, support and incentives for curation are available. Since that time, ubiquitous references to long tail data in the literature characterize this distribution in terms of size of data, demand for data, visibility and accessibility of data, and level of research competition with respect to data (Borgman, et al., 2016; Brooks, et al., 2016; Ferguson, et al., 2014; Liang, et al., 2010; Malik & Foster, 2012; Palmer, et al., 2007; Wallis, Rolando & Borgman, 2013). The ubiquitousness with which this terminology has been deployed indicates that data management issues abound across disciplines, despite the existence of data management plans, disciplinary infrastructures, improved standards and increased awareness of the benefits of data sharing for individual researchers and research communities.

Rapid changes in science practices towards open science, team science, and reproducibility – highlighted by the widespread adoption of the FAIR principles for making data Findable, Accessible, Interoperable and Reproducible (Wilkinson, et al., 2016) – lend urgency to conversations about what to do with typically-heterogeneous long tail data across disciplines. As demonstrated in previous studies (Heidorn, Stahlman & Steffen, 2018; Stahlman, 2020; Cragin, et al., 2010; Tenopir, et al., 2011; Gallaher, et al., 2015), much long tail data are "dark data" that remain hidden from other researchers for reasons such as lack of time and funding for data management, as well as lack of appropriate infrastructures, incentives, norms and standards, in addition to reasons related to collaboration and intellectual property, and worries about being "scooped". Increasingly, researchers combine data from multiple sources, using software and hardware tools to work with large datasets more easily. The size of both data and research teams are growing (NASEM, 2018a, 2018b), increasing the need for data management research and infrastructures to assist in assuring that all relevant data products associated with a research project and associated publications are shared. To begin tackling the issue of dark data in the long tail on an institutional and cultural scale, it is necessary to better understand where and how these data are produced, and the exploratory study presented here was overall motivated by the following questions:

>  Where are the resources for research allocated?
>  What is the relative distribution of funds and research topics?

What are the emerging research fronts that will require nuanced data management infrastructures in the future to prevent dark data?
Could a better understanding of the allocation of resources inform data management initiatives?

**Related Work**

This study aims to further develop a model and theoretical framework for understanding and successfully managing long tail data, and also to inform research and funding policies and curation efforts, where differences in type and topic of research and funding may impact data management practices. To work towards these objectives, we employ LDA topic modeling as a method to explore the ecosystem of research funding. LDA is a statistical model for analyzing document similarity (Blei, Ng & Jordan, 2003), which classifies documents by examining probabilistic associations between individual words and documents in a corpus, and has been widely used to derive insight from textual data in information science research (Figuerola, Marco & Pinto, 2017; Wolfram, 2016; Sugimoto, et al., 2011).

Similar to our study, Shi, et al. (2010) examine NSF grant proposal abstracts and associated publications using LDA, to explore temporal relations between funding and scientific output, and to inform funding policies. The authors focus on lag time between grants and publications in Computer Science, showing that in some cases research on certain topics is published several years before a related NSF grant is awarded, while for other topics NSF grants precede associated publications. Also informing the present study, which uses astronomy as a case study for the purpose of methodological development, Stahlman (2020) implemented a recent 2019 survey of authors of astronomy journal articles, which obtained detailed information about the characteristics and locations of underlying data for 211 papers published between 1994 and 2019. This study found that nearly three-quarters of research papers correspond to some "dark data" that are not accessible by other researchers, and that types of dark data vary over time since paper publication. The survey also found that certain characteristics of researchers, papers and data can be indicators of instances of underlying dark data, including advanced career stage of authors, higher number of authors on a paper, and datasets combined from multiple sources. Finally, the survey did not detect an explicit association between dark data and funding in astronomy through analysis of papers published over time.

**Research Design, Methodology and Results**

Through the present study and planned future studies, we contribute to further development of a "long tail" theoretical framework for research and data practices, while illuminating considerations for funders and curators, and presenting the study as a framework for deeper understanding of research and data lifecycles (Huang, Lee & Palmer, 2020; Borgman, 2019).

This paper demonstrates LDA as a method of illuminating long tail dynamics in publicly funded research by mining the text of National Science Foundation funding proposal abstracts and drawing insight from journal publications associated with funded grant proposals. The study design was guided by the following research questions:

> **RQ1**: Which topics have relatively more and fewer resources?
> **RQ2**: Is financial investment in a topic related to research output?
> **RQ3**: Which types of data likely correspond to topics in the head and tail of the funding distribution?

*Data Collection*

NSF proposals in the Division of Astronomical Sciences (AST) originating in 2016 were selected for this study (n=201). For exploratory insight and methodological testing, this relatively small dataset of NSF grant proposal abstracts and associated metadata was downloaded from NSF's awards database. We chose to focus on grants originating in 2016 because these grants are coming to fruition now and have had the same amount of time to produce published research outcomes. In many cases, collaborative grants are approved with multiple grant proposals sharing the same title and abstract; in these cases, we combined collaborative proposals into a single record and aggregated the funding amounts and program codes.

In parallel, we queried the Web of Science (WoS) Core Collection for records of published journal articles associated with our sample of NSF proposals (n=700, as of April 25, 2020). Funding information in WoS is fairly comprehensive for the time period of interest (2016-2020): of 92,479 articles in WoS within the Astronomy & Astrophysics subject heading, approximately 85% of records in this subject area contain funding information.

*Methods*

LDA topic analysis of grant proposal abstracts (n=201) was conducted. The text of the abstracts was preprocessed in R using standard natural language processing techniques such as tokenization, lemmatization and stopword removal. Other normalization measures include combining collaborative grants into single documents, as well as integrating the associated funding amounts and program codes. A log-likelihood method (Griffiths & Steyvers, 2004) was used to determine the ideal number of 22 topics for the corpus. Some topics have high affinity for larger grants, while some are associated with lower funding and some have a broad mix of funding level.

To illuminate the distribution of financial resources across topics, we introduce a new "Topic Investment" measure. Topic Investment (TI) establishes an estimate of the relative financial value of each topic in an analysis, though it says nothing about the intellectual merit of the topics. Topics are composed of sets of documents - in this case NSF grant proposal abstracts, where each grant has a particular affinity to the topic (LDA gamma) and a certain number of Dollars Awarded (DA) to each grant. The TI equation below distributes the DA across topics based on the relative LDA gamma values. The sum of the accumulated Topic Investments (TI) for all topics is equal to the sum of the Dollars Awarded (DA) for all grants in the period being studied (one year) or the total Annual Program Spending (APS). The sum of the TI for the number of topics (nt) also equals the APS. The Number of Grants (ng) is the number of grants for the study period. Across topics, when calculating TI we apportion a fraction of the DA proportional to the gamma across topics. If the grant has a high gamma in one topic and a low gamma in another topic more dollar worth is assigned to the topic with the higher gamma and less is assigned to the topic with lower gamma. Likewise, within each topic, all gamma values in a topic i sum to 1. To calculate TI we sum the product of the Dollar Amount of each grant (DAj) to the topic times the gamma for that Topic i (GammaTi).

$$\text{APS} = \sum_{i=1}^{ng} DA = \sum_{1=1}^{nt} TWi$$

$$TWi = \sum_{j=1}^{ng} DAj * GammaTi$$

Following the initial identification of 22 topics, the 5 proposal abstracts with highest gamma values for each topic were examined in depth alongside our dataset of associated published journal articles. Qualitative content analysis of these "top 5" proposal abstracts facilitated loose interpretation of each topic, summarized in Table 1 below. Tests for statistical significance were also conducted with variables related to award funding amount and research output, and the qualitative review of the top 5 proposal abstracts was used to classify the primary objective of each proposal and to further illustrate composition of topics in terms of research output. Although proposals can communicate multiple objectives, the main focus of each of the top 5 proposals was selected as the classifier (research; instrumentation; major facility construction and operation; education and training; or software and analysis techniques). For example, there is an education/training element for virtually every proposal, but if this was not the main focus of the proposal, a more appropriate classifier was selected. Proposals that do focus on education/training typically propose postdoctoral fellowships, along with projects such as the La Serena School for Data Science[1] and Research Experiences for Undergraduates (REU) training programs.

*Results*

---

[1] http://www.aura-o.aura-astronomy.org/winter_school/

Tests for statistical significance were conducted with variables related to award funding amount and research output, but no significant association was detected: for TI and number of articles the Pearson correlation coefficient is r =-.281 (p=.205), and for funding amount of individual grants and number of articles the Pearson correlation coefficient is r=-.071 (p=.472). Nevertheless, we observe in Table 1, Figure 1 and Figure 2 below that infrastructure and facilities grants occur in the top 20% of the funding distribution (Topics #1, #3, #16 and #21), and that topics that are heavily dominated by experimental research (Topics #2, #5, #7, #11, #14, #17, #18, and #22) produce more papers relative to topics dominated by areas such as facilities, instrumentation and education.

| Topic number | Topic label (based on bigrams and content analysis) | Number of publications with a top 5 grant acknowledged in WoS | Topic Investment Amount (in US Dollars) | Topic Investment Rank (1=highest investment) | Composition of top 5 grants associated with each topic |
|---|---|---|---|---|---|
| 1 | Astronomical instrumentation | 15 | 15849351 | 3 | 60%F; 20%R; 20%I |
| 2 | Cosmology | 93 | 7175474 | 17 | 100%R |
| 3 | Observatory management | 3 | 238285666 | 1 | 100%F |
| 4 | Radio astronomy | 34 | 8252839 | 12 | 20%I; 60%ST; 20%R |
| 5 | Active galaxies | 39 | 6445119 | 21 | 100%R |
| 6 | Exoplanets - technology | 7 | 11793068 | 6 | 100%I |
| 7 | Planet formation | 11 | 6808247 | 19 | 100%R |
| 8 | Techniques and training | 37 | 7556216 | 16 | 80%E; 20%R |
| 9 | Epoch of Reinoization | 43 | 10363471 | 8 | 40%R; 40%I; 20%E |
| 10 | Black holes | 21 | 7740729 | 14 | 80%R; 20%E |
| 11 | Planetary systems | 19 | 7694161 | 15 | 100%R |
| 12 | Terrestrial atmospheres | 9 | 6463311 | 20 | 80%R; 20%I |
| 13 | Dark matter | 26 | 8228472 | 13 | 80%R; 20%ST |
| 14 | Supernovae | 35 | 7145564 | 18 | 100%R |
| 15 | Data science | 26 | 5793256 | 22 | 60%ST; 20%R; 20%E |
| 16 | Astronomy training | 11 | 14034118 | 4 | 100%E |
| 17 | Stellar evolution | 21 | 8527815 | 10 | 100%R |
| 18 | Galaxy formation | 52 | 10661390 | 7 | 100%R |
| 19 | Exoplanets - research | 46 | 8488670 | 11 | 80%R; 20%ST |

| 20 | Dark energy | 115 | 13531286 | 5 | 60%ST; 20%F; 20%R |
| 21 | Large facilities | 1 | 59085683 | 2 | 60%F; 20%I; 20%E |
| 22 | Solar system | 45 | 8850868 | 9 | 100%R |

Table 1: Characterization of topics identified by LDA.
R = Research; I = Instrumentation; F = Major facility construction/operation;
E = Education/training; ST = Software/analysis techniques

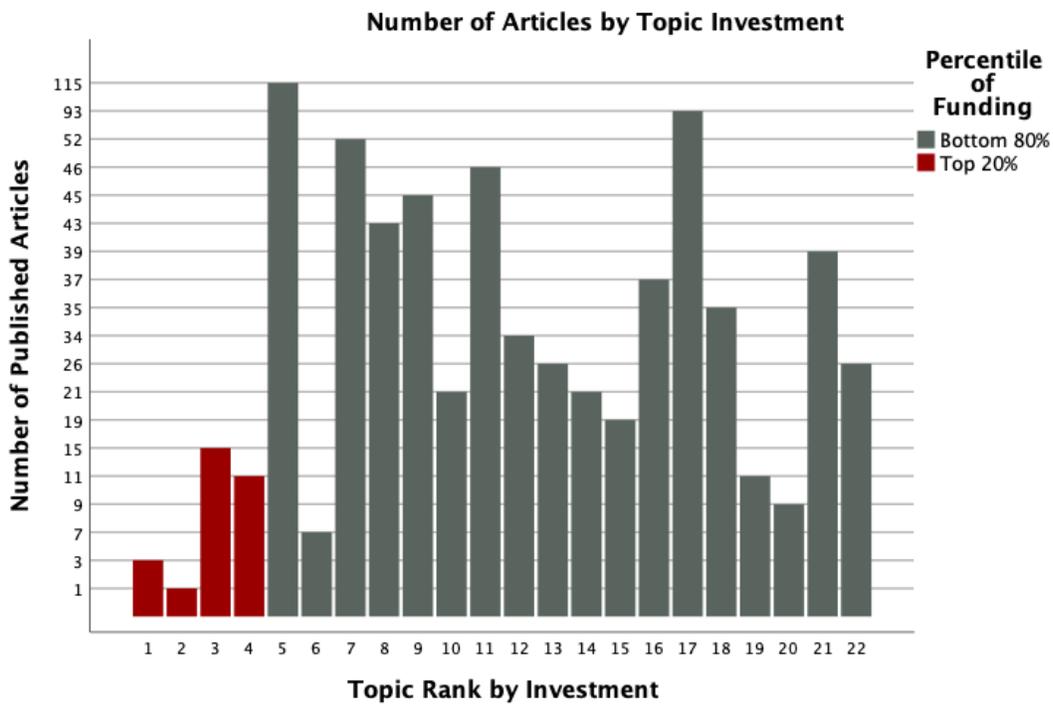

Figure 1: Number of articles in WoS for each topic by investment score

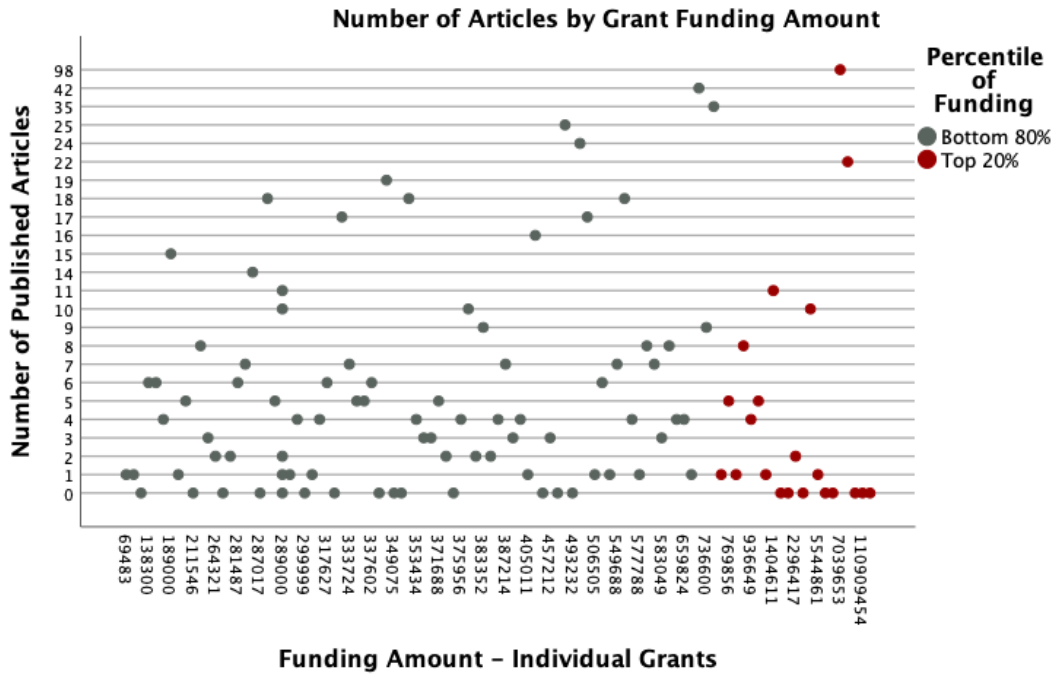

**Figure 2: Number of articles in WoS for each individual grant**

**Discussion**

This study has inquired about: the distribution of resources among research topics; whether topic investment is associated with research output; and which types of data likely correspond to topics in the head and tail of NSF's long tail funding distribution. Ultimately, we are interested in the distribution and accessibility of research data, and how services can be developed with deeper knowledge of dynamics of topics of funded research. While the present study used number of publications as a proxy for data and did not directly examine research data associated with funded grant proposals, the current study and methodology represent a first step in this direction.

Surprisingly, we did not detect a direct statistical association between funding amount and research output ascertained by number of publications for topics in astronomy. However, through examination of the top 5 proposal abstracts associated with each topic - alongside resulting publications as tangible research output - some interesting and relevant observations have surfaced that challenge and inform our initial model for a long tail distribution of funding and research output. Within the context of the findings of Stahlman (2020) that approximately three-quarters of published articles in astronomy result in inaccessible "dark data", the observations of the present study with respect to research output may be used to infer the risk and reality of dark data across funded research topics in astronomy.

If we consider the top 20% of the funding distribution as the "head", topics #1, #3, #16, and #21 clearly demonstrate the dynamics predicted by the long tail theoretical framework. The topic with the highest TI value is #3, which encompasses observatory management, while not demonstrating high publication output. Similarly, topic #21 demonstrates large facility management, while including instrumentation and education components and not publications. Topic #1 focuses on instrumentation, while including some research and instrumentation and 15 associated published journal articles. Topic #16 is entirely focused on education/training but has 11 associated published journal articles. In other words, highly funded topics are typically comprised of infrastructure grants, where publications may come later through research utilizing the facilities; these dynamics will be explored in future research.

Considering the bottom 80% of the funding distribution as "tail" topics, these topics also largely demonstrate the dynamics predicted by the long tail theoretical framework, with some intriguing variations. For example, topic #20 is technically in the bottom 80% with ~$135 million funding, but with a large number of associated papers (115). This topic corresponds to dark energy research, for which frequent papers are published to communicate data releases of the Dark Energy Survey[2] to the community, resulting in many more published papers by authors utilizing the data. Topic #7 "Planetary formation" is 100% research-oriented, but with few associated published papers in WoS, perhaps in part because of the cutting-edge and competitive research conducted by planetary scientists, as this community awaits and prepares for the upcoming launch of the James Webb Space Telescope[3] for exoplanet research. Conversely, topic #8 is 100% education/training-oriented, with 4 out of 5 top documents describing postdoctoral fellowships and a healthy number of published papers.

Overall, the long tail theory as it has been framed thus far does not apply seamlessly to the discipline of astronomy, which relies on global collaborations and very large investments in sophisticated instrumentation for collecting, disseminating and analyzing data. Our analysis demonstrates the unique social characteristics of astronomy research as well as the delicate ecosystem of funding between support for research and development of instruments and infrastructure. However, our analysis also supports aspects of the long tail theoretical framework. Particularly, considering that Stahlman (2020) illuminates the prevalence of "dark data" in astronomy, the topics within our study that correspond to relatively more publications may demonstrate more risk for producing dark data through the publication process, where derivative data products are often generated during analysis.

---

[2] https://www.darkenergysurvey.org/
[3] https://www.jwst.nasa.gov/

**Conclusion**

This paper overall shows how funding is directed to satisfy the research priorities of a certain scientific community (astronomy in this case) and informs further efforts to explore and develop targeted data curation resources and initiatives. Ongoing and future research will enhance the analysis presented here by: incorporating analysis of NSF program codes; conducting a similar analysis on a much larger dataset across disciplines and over time; and incorporating LDA topic analysis of associated publications as well. We will also expand our content analysis to look more deeply into research output and data behind papers associated with funded grants. The top 5 grants associated with a topic are an imperfect representation of each topic, and we intend to build upon this strategy by using the gamma measure to weight publication output for each topic. We may also automate classification of proposal abstracts using human annotation and machine learning for larger-scale insight.

For future research, it is fundamentally interesting to demonstrate disciplinary differences and similarities with respect to the allocation of research funding and implications for data across disciplines, where cross-disciplinary and convergent communication is facilitated by new infrastructures and shared methods. Especially considering the work of Shi, et al. (2010), the topics we have demonstrated here may represent research fronts that NSF has already endorsed and areas that NSF has yet been hesitant to fully support. In the case of endorsement of research fronts, our analysis may point to areas that will bridge subdisciplines and further draw coherent boundaries for the sciences (Varga, 2019). In conclusion, the study presented here has laid the groundwork for extensive exploration across funding agencies and disciplines, and seeking to link topics with data to further inform a theoretical framework for long tail data.